# Photo-nucleation theory of correlation of Stream-flow of four South American rivers with Sunspot Cycle

W Byers Brown[1]

Department of Chemistry
University of Manchester
Oxford Road, Manchester M13 9PL, UK

## Abstract

C.T.R.Wilson showed that when supersaturated water vapour was exposed to ultraviolet radiation from sunlight or other sources in the presence of oxygen it immediately condensed to form an aerosol. This phenomenon was eventually explained as due to the formation of a charge-transfer complex $H_2O^+O_2^-$, whose existence was confirmed theoretically and subsequently established experimentally. It is proposed that the correlation recently discovered between the stream-flow of the Parana and three other rivers in South America and the solar sunspot cycle is due to the photo-nucleation mechanism investigated by Wilson.

[1]billbyersbrown@me.com

## 1 Introduction

Thanks to the recent work by Pablo J.D.Mauas and collaborators an extraordinary correlation has been discovered between the stream-flows of the Parana [1] and three other South American rivers [2] and the solar sunspot cycle. Their conclusion is that the higher solar activity corresponds to a larger precipitation both in summer and wintertime, and is probably due to increased solar irradiance.

In the absence of alternative explanations, it is suggested here that the increased precipitation could be due to photo-nucleation caused by enhanced ultra-violet radiation from sunspots.

**2 Photo-nucleation**

Over one hundred years ago, following earlier work by Lenard & Wolf [3], it was shown conclusively by C.T.R.Wilson that when supersaturated water vapour was exposed to ultraviolet radiation from sunlight or other sources in the presence of oxygen it immediately condensed to form an aerosol [4]. This example of photo-nucleation has been largely ignored by atmospheric scientists for over a century, despite being repeated by a number of other workers [5,6,7,8,9]. It remained unexplained until after the discovery of charge-transfer (CT) reactions and complexes in the second half of the 20$^{th}$ century. It was then suggested [10] that the photo-nucleation investigated by Wilson was due to the formation of the charge-transfer complex $H_2O^+O_2^-$, whose large dipole moment would attract water molecules and ultimately lead to water droplets. Quantum chemistry calculations [11] confirmed the stable existence of the charge-transfer complex with an excitation energy of 5.9 eV above the collision complex $H_2O \cdot O_2$, corresponding to a UV excitation wavelength of 210 nm, and a dipole moment of 6.2 Debye; this is more than adequate to account for the power of the electric field of a CT complex to attract water molecules (dipole moment 1.8 Debye) and create a cluster of water molecules. Direct experimental evidence for the existence of the charge-transfer complex was obtained by Cacace, de Petris, Pepi & Troiani [12] by means of neutralisation-reionization mass spectroscopy [13] using the 16 and 18 isotopes of oxygen.

## 3 Kinetic scheme for Wilson clusters

It is straightforward to sketch a kinetic scheme for the creation of Wilson clusters from water and oxygen [10]. We assume the van der Waals collision complex $H_2O.O_2$ (vdW) is in fast mobile equilibrium with oxygen and water,

$$H_2O + O_2 \rightleftharpoons H_2O.O_2, \qquad (1)$$

so that

$$[vdW] = K_V [O_2][H_2O], \qquad (2)$$

where $K_V$ is an equilibrium constant. The vdW complex can absorb a UV photon of frequency $\upsilon$ to form $H_2O^+O_2^-$, (CTC), which can be assumed to have a short radiative life-time $\tau$ in the gas phase,

$$H_2O.O_2 + h\upsilon \rightarrow H_2O^+O_2^- \rightarrow H_2O.O_2. \qquad (3)$$

The dominant terms in the rate equation for CTC are

$$d[CTC]/dt = (\sigma I/h\upsilon)[vdW] - [CTC]/\tau, \qquad (4)$$

where $I(\upsilon)$ is the UV intensity and $\sigma(\upsilon)$ is the absorption cross-section of vdW. The steady state concentration of CTC is therefore given by

$$[CTC] = (\tau\sigma I/h\upsilon)[vdW], \qquad (5)$$

and requires for its estimation values of $\sigma$, $\tau$, $\upsilon$, $I$ and $K_V$.

If we assume that all the $O_2$ molecules in oxygenated liquid water are weakly attached to $H_2O$ molecules to form $H_2O \cdot O_2$, then σ can be obtained from the photo-absorption results of Heidt and Johnson [14], who found for water saturated with oxygen σ(200 nm) = 17.1 $pm^2$ at 24° C.

If we assume that the vdW complex is very loose, so that the internal vibrations and rotations of $O_2$ and $H_2O$ are unaffected by complexing, and further assume they interact by an exponential-α:n intermolecular potential with binding energy $D_e$ at intermolecular distance $R_e$, then

$$K_V = (4\pi L R_e^3) \sqrt{\pi k T (\alpha - n)/\alpha n(\alpha - n - 1) D_e} \exp(D_e/kT), \qquad (6)$$

where k is the Boltzmann constant, and L is the Avogadro number. By substituting α=12 (for the exponential repulsion exponent) and n=4 (for long-range dipole-induced-dipole interaction) and the calculated values [11] for $D_e$ and $R_e$ we get

$$K_V = 5.3 \sqrt{T} \exp(433/T) \text{ cm}^3 \text{ mol}^{-1} \qquad (7)$$

where T is in °K. The radiative life-time τ of the CTC has been estimated from the neutralization-reionization mass spectroscopy results [12] to be of the order of a microsecond, so $\tau \approx 1 \mu s$.

The rate of production of the first Wilson cluster

$$H_2O^+ O_2^- + H_2O \rightarrow W_1, \qquad (8)$$

containing just one CTC can be estimated by simple collision theory, but in fact requires sophisticated molecular dynamics (MD) for accuracy, as indicated in references [15 & 16]. The CTCs may be lost by reacting with the water in the cluster to form, ultimately, hydrogen peroxide,

$$H_2O^+O_2^- + H_2O \rightarrow 2H_2O_2 . \qquad (9)$$

Otherwise, $W_1$ will go on to coalesce with other $H_2O$ molecules to form larger Wilson clusters. These later stages of clustering will require MD calculations of the kind used to describe the formation of pure water clusters.

In the upper atmosphere, clusters can be assumed to accumulate the number of $H_2O$ molecules necessary for equilibrium with the ambient vapour, or until they reach a critical nucleation number of the CTC or $H_2O_2$ stabilizing species for the given saturation ratio S at the given temperature T, and are able to grow to droplet size. The actual process of cloud formation and precipitation presumably parallels in complexity that for droplets formed round condensation nuclei [17].

**4 Discussion**

It is well-known that the total emission of UV by sunspots is greater at the maxima [17], but there does not appear to be any complete data about the emission spectrum, and in particular whether the intensity of shorter wave-length UV increases at the maxima, which appears probable from the visual increase of radiation from the rims (faculae or plages) of the hot spots. In any case, as Katz and co-workers have shown [9], photo-nucleation occurs up to 320 nm and beyond, albeit weakly, and this may be sufficient to lead to Wilson clusters and rainfall.

It should be noted that Wilson [4] was able to cause condensation of supersaturated water vapour using solar UV virtually at sea level (Cambridge).

**5 Conclusion**

The main reason for the neglect by atmospheric scientists of the photo-induced nucleation of water vapour in the presence of oxygen and the absence of pollutants, apart from ignorance of its existence, must be the lack hitherto of a simple explanation of what is going on at the molecular level, and therefore a means of estimating its importance. However, until the emission spectrum of UV during sunspot maxima is known it will not be possible to convincingly model the process [15,18] and assess the likelihood of the correlation of the streamflows of the four South American rivers with sunspot variation [1,2] being due to photo-nucleation.

There is no reason to suppose that if photo-nucleation is taking place in the upper atmosphere over South America it is limited to that region. It may be that globally there is a layer of the atmosphere where in the presence of UVC radiation from sunspots, which has not encountered the ozone layer, Wilson clusters can be formed and give rise to rainfall.

Further experiments with modern large cloud chambers and further development of the theory could provide invaluable confirmation of the approach presented here and will hopefully be forthcoming.

**Acknowledgements**

I am grateful to Dr Ann Webb, Dr Clive Saunders and Dr Paul Connolly of the School of Earth, Atmospheric and Environmental Sciences of the University of Manchester for helpful discussions of the above proposal. I am also grateful to Prof Jerrald Hague for his continuing interest and support for this project.